\documentstyle[twocolumn,prb,aps,graphicx,floats,amssymb]{revtex}
\draft
\begin{document}

\twocolumn[
\hsize\textwidth\columnwidth\hsize\csname@twocolumnfalse\endcsname
\draft

\title{Integer quantum Hall transition in the presence of a
  long-range-correlated quenched disorder}

\author{P. Cain, R. A.\ R\"omer, and M. Schreiber}
\address{Institut f\"ur Physik, Technische Universit\"at Chemnitz,
  D-09107 Chemnitz, Germany}
\author{M. E.\ Raikh}
\address{Department of Physics, University of Utah, Salt Lake City, Utah
  84112}

\date{$Revision: 1.39 $, compiled \today} \maketitle

\begin{abstract}
  We theoretically study the effect of long-ranged inhomogeneities on
  the critical properties of the integer quantum Hall transition.  For
  this purpose we employ the real-space renormalization-group (RG)
  approach to the network model of the transition.  We start by
  testing the accuracy of the RG approach in the absence of
  inhomogeneities, and infer the correlation length exponent $\nu=
  2.39$ from a broad conductance distribution.  We then incorporate
  macroscopic inhomogeneities into the RG procedure.  Inhomogeneities
  are modeled by a smooth random potential with a correlator which
  falls off with distance as a {\em power law}, $r^{-\alpha}$. Similar
  to the classical percolation, we observe an enhancement of $\nu$
  with decreasing $\alpha$.  Although the attainable system sizes are
  large, they do not allow one to unambiguously identify a cusp in the
  $\nu(\alpha)$ dependence at $\alpha_{c}=2/\nu$, as might be expected
  from the extended Harris criterion.  We argue that the fundamental
  obstacle for the numerical detection of a cusp in the {\em quantum}
  percolation is the implicit randomness in the {\em Aharonov-Bohm
    phases} of the wave functions. This randomness emulates the
  presence of a {\em short-range} disorder alongside the smooth
  potential.
\end{abstract}

] 
\narrowtext
\tighten

\section{Introduction}
\label{sec-intro}

The critical behavior of electron wave functions in the vicinity of
the integer quantum Hall (QH) transition is now well
understood.\cite{Huc95} That is, the localization length diverges as
$\varepsilon^{-\nu}$, where $\varepsilon$ is the deviation from the
critical energy. The most accurate value of the exponent $\nu$
extracted from numerical simulations is $\nu=2.35\pm0.03$.\cite{HucK90}
 On the experimental side, the study of the critical
behavior of the resistance in the transition region at strong magnetic
field $B$ has a long history which can be conventionally divided into
three periods.

\paragraph*{\rm (a)} The first experimental works
\cite{WeiTP85,WeiTPP88,KocHKP91a,KocHKP91b,KocHKP92,WeiLTP92,HwaWET93}
reported a narrowing of the transition peak with temperature $T$, as
$T^\kappa$ with $\kappa\sim 0.4$. The spread in the actual value of
$\kappa\propto 1/\nu$ measured in different experiments was attributed
to the difference in the type of disorder in the samples of Refs.\ 
\onlinecite{WeiTP85,WeiTPP88} and \onlinecite{KocHKP91a}.  Another
experimental method to explore the critical behavior was employed in
Refs.\ \onlinecite{KocHKP91b} and \onlinecite{KocHKP92}, where
$\kappa$ was deduced from the sample size dependence of the width,
$\Delta B$, of the transition region. The value of $\kappa$ obtained
by this technique appeared to be consistent with temperature
measurements of Ref.\ \onlinecite{KocHKP91a}, in the sense that
$\kappa$ was found to be sample dependent. On the other hand, it was
argued in Ref.\ \onlinecite{WeiLTP92} that the lack of universality in
Refs.\ \onlinecite{KocHKP91a,KocHKP91b,KocHKP92} has its origin in the
long-ranged character of the disorder in GaAs-based heterostructures
studied in these works. This is because for a smooth disorder the
energy interval within which the electron transport is dominated by
localization effects is relatively narrow.\cite{WeiLTP92} The
measurements in Refs.\ \onlinecite{WeiTP85} and \onlinecite{WeiTPP88}
suggesting the universality of $\kappa$ were carried out on
In$_x$Ga$_{1-x}$As/InP heterostructures in which disorder is believed
to be short-ranged.\cite{Col99} Despite the disagreement about
universality of the exponent $\kappa$, the fact that the narrowing of
the plateau transition occurs as $T^\kappa$ was not questioned in
Refs.\ 
\onlinecite{WeiTP85,WeiTPP88,KocHKP91a,KocHKP91b,KocHKP92,WeiLTP92,HwaWET93}.

\paragraph*{\rm (b)}
The absence of scaling was reported first for the QH-insulator
transition \cite{ShaHLT98} and then for the plateau-plateau
transition.\cite{BalMB98} In the latter paper the conclusion about the
absence of scaling was drawn from the analysis of the frequency
dependence of $\Delta B$ in GaAs/Al$_y$Ga$_{1-y}$As heterostructures
(in contrast to the similar analysis in Ref.\ \onlinecite{EngSKT93}).
That is, the authors of Refs.\ \onlinecite{ShaHLT98} and
\onlinecite{BalMB98} concluded, that the width of the transition
region {\em saturates} as $T\rightarrow 0$. A possible explanation of
this behavior \cite{ShiAK98,Shi99} is based on the scenario of
tunneling between electron puddles with a size larger than the
dephasing length. The microscopic origin of these puddles was
attributed to sample inhomogeneities.\cite{SimH94,RuzCH96,CooHHR97}

\paragraph*{\rm (c)} Very recent experimental results \cite{SchVOW00} on
scaling of plateau-insulator as well as plateau-plateau QH transitions
carried out on the same In$_x$Ga$_{1-x}$As/InP sample as in Ref.\
\onlinecite{HwaWET93} suggested that the narrowing of the transition
width with temperature follows a power-law dependence $\Delta B
\propto T^\kappa$ with $\kappa\approx 0.4$.  Even when the authors of
Ref.\ \onlinecite{SchVOW00} analyzed their data using the procedure of
Ref.\ \onlinecite{ShaHLT98}, i.e., by plotting the logarithm of the
longitudinal resistance versus $\Delta B$, they obtained straight
lines with slopes proportional to $T^{\kappa'}$ with $\kappa'\approx
0.55$. They attributed the difference between $\kappa$ and $\kappa'$
to the marginal dependence of the critical resistance on $T$. It was
also speculated in Ref.\ \onlinecite{SchVOW00} that this dependence
most likely results from macroscopic inhomogeneities in the sample.
In the latest papers \cite{KucMDK00,HohZHP00,HohZH00} the frequency
dependence of the QH transition width was studied. The results did not
support the saturation of the width,\cite{ShaHLT98,BalMB98} but
rather confirmed the scaling hypothesis.

Summarizing, it is now conclusively established that insulator-plateau
and plateau-plateau transitions exhibit the same critical behavior. It
is also recognized that macroscopic inhomogeneities can either mask
the scaling or alter the value of $\kappa$.\cite{SchVOW00}

On the theoretical side, in all previous considerations
inhomogeneities were incorporated into the theory through a spatial
variation of the {\em local}
resistivity.\cite{ShiAK98,Shi99,SimH94,RuzCH96,CooHHR97,inhomo} In
other words, all existing theories are either "homogeneous quantum
coherent" or "inhomogeneous {\em incoherent}". Meanwhile, there is
another scenario which has never been explored.  Close to the
transition the {\em quantum} localization length $\xi$ becomes
sufficiently large.  Then the long-ranged disorder can affect the
character of the divergence of $\xi$.  At this point we recall the
classical limit,\cite{WeiH83,Wei84} in which the long-ranged disorder
does affect the value of the critical exponent in the percolation
problem. Obviously, when the disorder is long-ranged but has a finite
correlation radius, one should not expect any changes in the critical
behavior. The principle finding of Refs.\ \onlinecite{WeiH83} and
\onlinecite{Wei84} is that the critical exponent can change when the
correlator of the disorder $\langle V({\bf r})V({\bf r'})\rangle$
falls off with distance as a power law, i.e., $\propto |{\bf r}-{\bf
  r'}|^{-\alpha}$, (quenched disorder). According to Refs.\ 
\onlinecite{WeiH83} and \onlinecite{Wei84} the critical exponent of
the classical percolation $\nu=4/3$ crosses over to $\nu=2/\alpha$ for
$\alpha<3/2$, i.e., when the decay of the correlator is slow enough.
In the present paper we study the effects of quenched disorder on {\em
  quantum} percolation.  The latter is known to describe the
localization-delocalization transition for a two-dimensional (2D)
electron in a strong magnetic field. As a model of quantum percolation
we employ the Chalker-Coddington (CC) model \cite{ChaC88} which is one
of the main "tools" for the quantitative study of the QH
transition.\cite{LeeWK93,LeeC94,WanLW94,LeeCK94,KagHA95,KagHA97,KleM95,KleM97,RuzF95,Met98b,JanMZ99,KleZ00}
The CC model is a strong magnetic field (chiral) limit of a general
network model, first introduced by Shapiro \cite{Sha82} and later
utilized for the study of localization-delocalization transitions
within different universality
classes.\cite{FreJM98,MerJH98,Jan98,FreJM99,KagHAC99,ChaRKH00} In
addition to describing the QH transition, the CC model applies to a
much broader class of critical phenomena since the correspondence
between the CC model and thermodynamic, field-theory and
Dirac-fermions models
\cite{Lee94,Zir94,Zir97,GruRS97,LudFSG94,Kim96,HoC96,KonM96,MarT99}
was demonstrated.

In order to study the role of the quenched disorder on the
localization-delocalization transition we treat the CC model within
the real-space renormalization group (RG) approach.  First, in
Sec.\ \ref{sec-RG} we check the accuracy of the RG approach, and show that it
provides a remarkably accurate description of the QH transition.  In
Sec.\ \ref{sec-inh}  we extend the RG approach to incorporate the
quenched disorder. Concluding remarks are presented in Sec.\
\ref{sec-sum}.

\section{Test of the RG approach to the CC model}
\label{sec-RG}
%
\subsection{Description of the RG approach}
\label{ssec-RG-RG}
\begin{figure}
\centerline{\includegraphics[width=0.95\columnwidth]{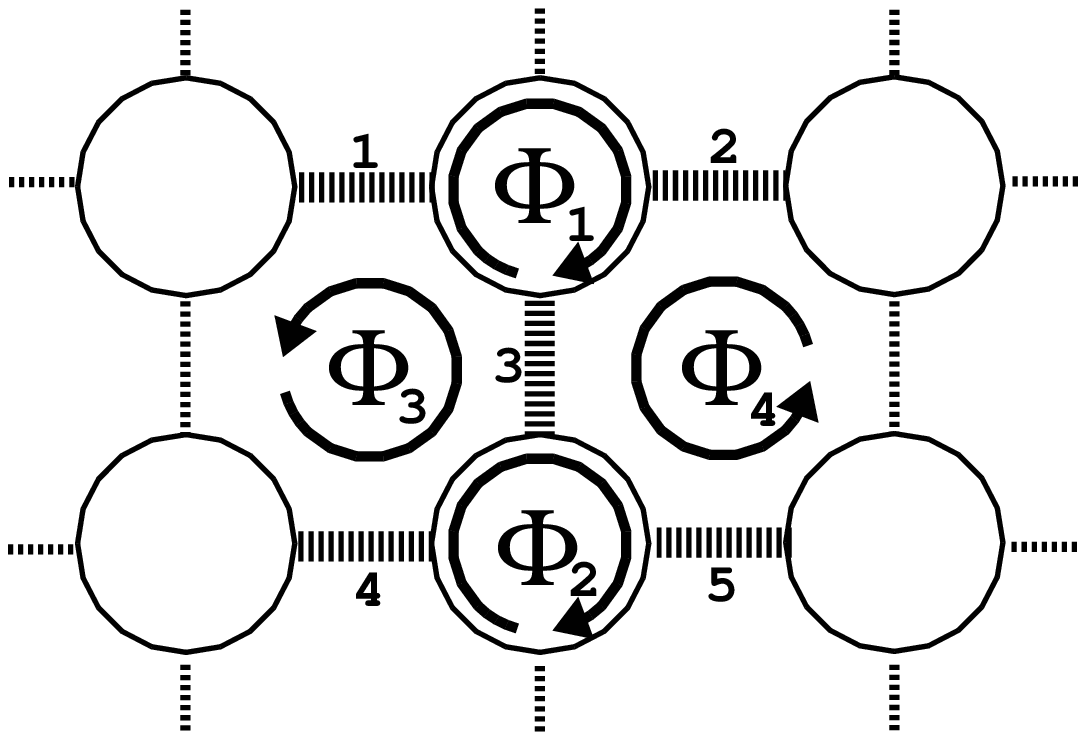}}
\caption{\label{fig-RGstruct}
  Network of SP's (dashed lines) and equipotential lines (circles) on
  a square lattice. The RG unit used for Eq.\ (\ref{eq-qhrg}) combines
  five SP's (numbered thick dashed lines) --- in analogy with
  classical 2D bond percolation RG (Refs.\ \protect\onlinecite{ReyKS77} and
  \protect\onlinecite{Ber78}) --- into a super-SP.  $\Phi_1, \ldots,
\Phi_4$ are the phases acquired by an electron drifting along the
contours indicated by the arrows.}
\end{figure}

As mentioned in Sec.\ \ref{sec-intro}, the CC model is a chiral
limit of the general network model.\cite{Sha82}  However, it was
originally derived from a microscopic picture of electron motion in
a strong magnetic field and a smooth potential.\cite{ChaC88} Within
this picture, the links can be identified with semiclassical
trajectories of the guiding centers of the cyclotron orbit, while the
nodes correspond to the saddle points (SP's) at which different
trajectories come closer than the Larmour radius.  For simplicity, the
nodes were placed on a square lattice.  We will use an equivalent, but
slightly different graphical representation of the CC network shown in
Fig.\ \ref{fig-RGstruct}.  In this representation the centers of the
trajectories of the guiding centers are placed on a square lattice and
play the role of nodes, whereas the SP's should be identified with
links.

We now apply a real-space RG approach \cite{GalR97,AroJS97} to the CC
network.\cite{SinMG00,ZulS01} The RG approach is based on the
assumption that a certain part of the network containing several SP's,
the RG unit, represents the entire network. This unit is then replaced
through the RG transformation by a single {\em super}-SP with an
$S$ matrix determined by the $S$ matrices of the constituting SP's.
The network of super-SP's is then treated in the same way as the
original network.  Successive repetition of the RG transformation
yields the information about the $S$ matrix of very large samples,
since, after each RG step, the effective sample size grows by a
certain factor determined by the geometry of the original RG unit.
Obviously, a single RG unit is a rather crude approximation of the
network.  Therefore, prior to applying the RG approach to the study of
the quantum Hall transition in the presence of quenched disorder, we
first check the accuracy of this approach for the conventional,
uncorrelated case, where the comparison with the results of direct
numerical simulations is possible.

The RG unit we use is extracted from a CC network on a regular 2D
square lattice as shown in Fig.\ \ref{fig-RGstruct}.  A super-SP
consists of five original SP's by analogy to the RG unit employed for
the 2D bond percolation problem.\cite{StaA92,ReyKS77,Ber78}
As in any RG scheme, the unit shown in Fig.\ \ref{fig-RGstruct} leaves
out a number of bonds of the original lattice. Nevertheless, it is well
known that the application of this scheme to the classical case yields
very accurate results.\cite{StaA92}

Between the SP's an electron travels along equipotential lines, and
accumulates a certain Aharonov-Bohm phase. Different phases are
uncorrelated, which reflects the randomness of the original potential
landscape.  Each SP can be described by two equations relating the
wave-function amplitudes in incoming and outgoing channels.  This
results in a system of ten linear equations, the solution of which
yields the following expression for the transmission coefficient of
the super-SP (Ref.\ \onlinecite{GalR97}):

\twocolumn[
\hsize\textwidth\columnwidth\hsize\csname@twocolumnfalse\endcsname
\draft
\begin{equation}
  \label{eq-qhrg}
  t'= \left | \frac{
  t_1 t_5 (r_2 r_3 r_4 e^{\imath\Phi_2} - 1) +
  t_2 t_4 e^{\imath (\Phi_3+\Phi_4)} (r_1 r_3 r_5 e^{-\imath\Phi_1} - 1) +
  t_3 (t_2 t_5 e^{\imath\Phi_3} + t_1 t_4 e^{\imath\Phi_4})
}
{
  (r_3 - r_2 r_4 e^{\imath\Phi_2}) (r_3 - r_1 r_5 e^{\imath \Phi_1}) +
  (t_3 - t_4 t_5 e^{\imath\Phi_4}) (t_3 - t_1 t_2 e^{\imath\Phi_3})
}\right | \quad .
\end{equation}
] 
\noindent Here $t_i$ and
$r_i=(1-t_i^2)^{1/2}$ are, respectively, the transmission and
reflection coefficients of the constituting SP's; $\Phi_j$ are the
phases accumulated along the closed loops (see Fig.\
\ref{fig-RGstruct}).  Equation (\ref{eq-qhrg}) is the RG transformation,
which allows one to generate (after averaging over $\Phi_j$) the
distribution $P(t')$ of the transmission coefficients of super-SP's
using the distribution $P(t)$ of the transmission coefficients of the
original SP's. Since the transmission coefficients of the original SP's
depend on the electron energy $\varepsilon$, the fact that
delocalization occurs at $\varepsilon = 0$ implies that a certain
distribution, $P_{\text c}(t)$ --- with $P_{\rm c}(t^2)$ being
symmetric with respect to $t^2=\frac{1}{{2}}$ --- is the fixed point
(FP) of the RG transformation Eq.\ (\ref{eq-qhrg}).  The distribution
$P_{\text c}(G)$ of the dimensionless conductance $G$ can be
obtained from the relation $G=t^2$, so that $P_{\text c}(G)\equiv
P_{\text c}(t)/2t$.

\subsection {Critical exponent within the RG approach}
\label{ssec-RG-Q}

Since the dimensionless SP height $z_i$ and the transmission
coefficient $t_i$ at $\varepsilon=0$ are related as
$t_i=(e^{z_i}+1)^{-1/2}$, transformation (\ref{eq-qhrg})
determines the height of a super-SP through the heights of the five
constituting SP's. Correspondingly, the distribution $P(G)$ determines
the distribution $Q(z)$ of the SP heights via $Q(z) = P(G) (dG/dz) =
\frac{1}{4}\cosh^{-2}(z/2) P\left[(e^z+1)^{-1}\right]$.
In fact, $Q(z)$ is not a characteristic of the actual SP's, but
rather, as we will see below, represents a convenient parametrization
of the conductance distribution.

The language of the SP heights provides a natural way to extract the
critical exponent $\nu$.  Suppose that the RG procedure starts with an
initial distribution, $Q_0(z)=Q_c(z-z_0)$, that is shifted from the
critical distribution, $Q_c(z)$, by a small $z_0\propto \varepsilon$.
Since $z_0\ll 1$, the first RG step would yield $Q_c(z-\tau z_0)$ with
some number $\tau$ independent of $z_0$. After $n$ steps the center of
the distribution will be shifted by $z_{{\rm max},n}=\tau^n z_0$, while
the sample size will be magnified by $2^n$. At the $n$th step
corresponding to $z_{{\rm max},n}\sim 1$ a typical SP is no longer
transmittable.  Then the localization length $\xi$ should be
identified with $2^n \propto z_0^{-\nu} \propto \varepsilon ^{-\nu}$,
with $\nu=\ln 2/\ln\tau$.  When the RG procedure is carried out
numerically, one should check that $z_0$ is small enough so that
$z_{{\rm max},n}\propto z_0$ for large enough $n$. Consequently, the
working formula for the critical exponent can be presented as
\begin{equation}
\label{eq-nu}
 \nu = \frac{\ln 2^n}{\ln \left(\frac{z_{{\rm max},n}}{z_0}\right)}
\end{equation}
which should be independent of $n$ for large $n$.

\subsection{Numerical results}
\label{ssec-RG-res}

In order to find the FP conductance distribution $P_{\text c}(G)$, we
start from a certain initial distribution of transmission
coefficients, $P_0(t)$ (see below).  The distribution is discretized
in at least $1000$ bins, such that the bin width is typically $0.001$
for the interval $t\in [0, 1]$. From $P_0(t)$, we obtain $t_i$,
$i=1,\ldots,5$, and substitute them into the RG transformation
[Eq. (\ref{eq-qhrg})].  The phases $\Phi_j$, $j= 1,\ldots,4$ are chosen
randomly from the interval $\Phi_j \in [0, 2\pi]$.  In this way we
calculate at least $10^{7}$ super-transmission coefficients $t'$.  The
obtained histogram $P_1(t')$ is then smoothed using a Savitzky-Golay
filter \cite{PreFTV92} in order to decrease statistical fluctuations.
At the next step we repeat the procedure using $P_1$ as an initial
distribution.  We assume that the iteration process has converged when
the mean-square deviation $\int{dt [ P_n(t)-P_{n-1}(t) ]^2}$ of the
distribution $P_n$ and its predecessor $P_{n-1}$ deviate by less than
$10^{-4}$.

We are now able to study samples with short-ranged disorder.  The
actual initial distributions, $P_0(t)$, were chosen in such a way that
corresponding conductance distributions, $P_0(G)$, were either uniform
or parabolic, or identical to the FP distribution found
semianalytically in Ref.\ \onlinecite{GalR97}.  All these
distributions are symmetric with respect to $G=0.5$. We find that,
regardless of the choice of the initial distribution, after $5$--$10$
steps the RG procedure converges to the {\em same} FP distribution
which remains unchanged for another $4$--$6$ RG steps.  Small
deviations from symmetry about $G=0.5$ finally accumulate due to
numerical instabilities in the RG procedure, so that typically after
$15$--$20$ iterations the distribution becomes unstable and flows
towards one of the classical FP's $P(G)=\delta(G)$ or
$P(G)=\delta(G-1)$.  We note that the FP distribution can be
stabilized by forcing $P_n(G)$ to be symmetric with respect to $G=0.5$
in the course of the RG procedure.

Figure \ref{fig-PQ} illustrates the RG evolution of $P(G)$ and $Q(z)$.
In order to reduce statistical fluctuations we average the FP
distributions obtained from different $P_0(G)$'s.  The FP distribution
$P_{\text c}(G)$ exhibits a flat minimum around $G=0.5$, and sharp
peaks close to $G=0$ and $G=1$. It is symmetric with respect to
$G\approx 0.5$ with $\langle G \rangle = 0.498\pm0.004$, where the
error is the standard error of the mean of the obtained FP
distribution.  The FP distribution $Q_c(z)$ is close to Gaussian.

\begin{figure}[p]
\centerline{\includegraphics[width=0.95\columnwidth]{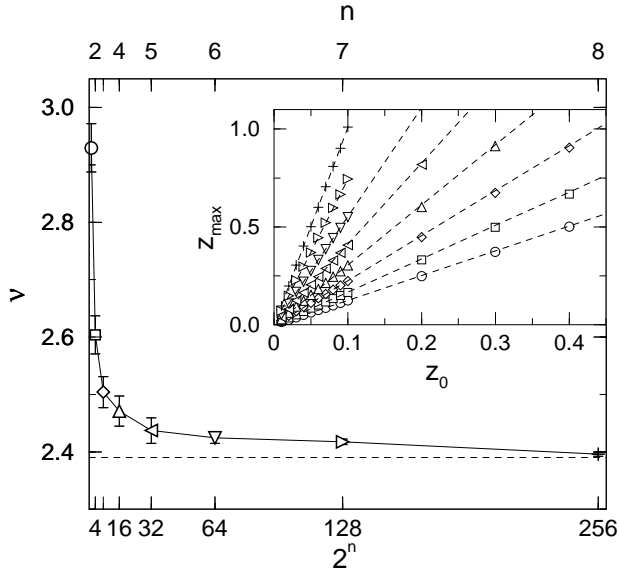}}
\caption{\label{fig-nu-L}
  Critical exponent $\nu$ obtained by the QH-RG approach as function
  of effective linear system size $L= 2^n$ for RG step $n$. The error
  bars correspond to the error of linear fits to the data.  The dashed
  line shows $\nu=2.39$. Inset: $\nu$ is determined by the dependence
  of the maximum $z_{{\rm max},n}$ of $Q_n(z)$ on a small initial
  shift $z_0$.  Dashed lines indicate the linear fits.}
\end{figure}

We now turn to the critical exponent $\nu$.  As a result of the general
instability of the FP distribution, an initial shift of $Q_c(z)$ by a
value $z_0$ results in the further drift of the maximum position,
$z_{{\rm max},n}$, away from $z=0$ after each RG step.  As expected,
$z_{{\rm max},n}$ depends linearly on $z_0$.  This dependence is shown
in Fig.\ \ref{fig-nu-L} (inset) for different $n$ from $1$ to $8$.
The critical exponent is then calculated from the slope according to
Eq.\ (\ref{eq-nu}).  Figure \ref{fig-nu-L} illustrates how the critical
exponent converges with $n$ to the value $2.39\pm0.01$.  The error
corresponds to a confidence interval of $95\%$ as obtained from the
fit to a linear behavior.

Due to the high accuracy of our calculation of $P_{\text c}(G)$, we
were able to reliably determine many central moments $\langle
(G-\langle G\rangle)^m\rangle$ of the FP distribution $P_{\text
  c}(G)$.  These moments are plotted in Fig.\ \ref{fig-pmom}.

\subsection{Comparison with previous simulations}
\label{ssec-RG-compare}

\begin{figure}
\centerline{\includegraphics[width=0.95\columnwidth]{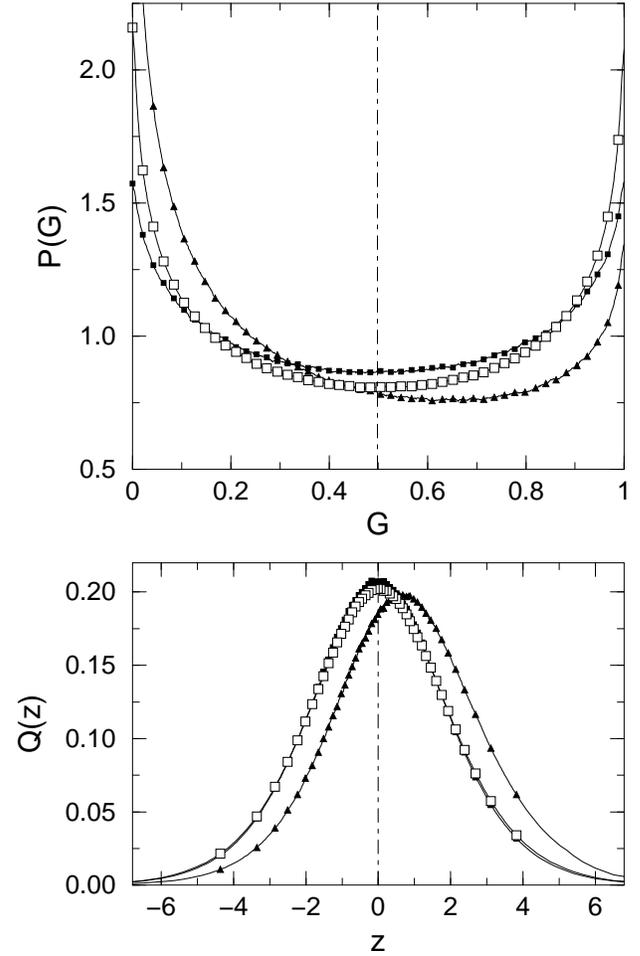}}
\caption{\label{fig-PQ} Top: $P(G)$ (thin lines) as function of
  conductance $G$ at a QH plateau-to-plateau transition.  Symbols mark
  every $20$th data point for the initial distribution
  ($\blacksquare$), the FP ($\Box$) and the distribution for RG step
  $n=16$ ($\blacktriangle$). The vertical dashed line indicates the
  average of the FP distribution.  Bottom: Corresponding plots for the
  distribution $Q(z)$ of SP heights.}
\end{figure}

\begin{figure}[p]
\centerline{\includegraphics[width=0.95\columnwidth]{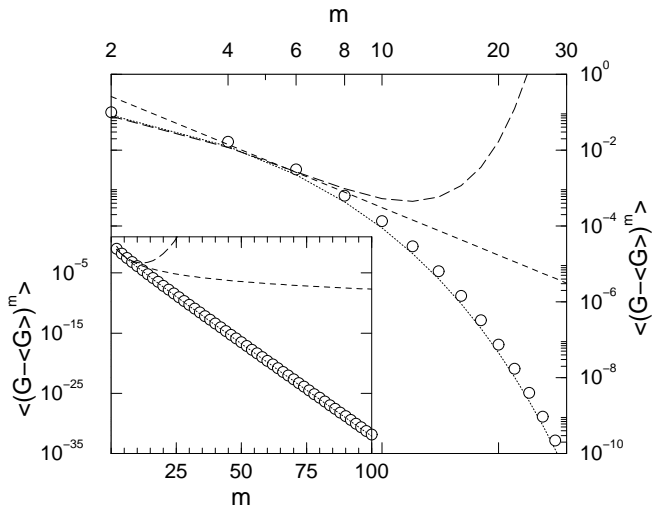}}
\caption{\label{fig-pmom}
  Moments $\langle (G-\langle G \rangle)^m \rangle$ of the FP
  distribution $P_{\text c}(G)$ ($\circ$). Dashed lines are results from
  Ref.\ \protect\onlinecite{WanJL96}. The dotted line corresponds to the
  moments of a constant distribution. Inset: Higher moments of $P_{\text
  c}(G)$ following an exponential behavior. The agreement of data and
  fit demonstrates the quality of the fit result.}
\end{figure}

By dividing the CC network into units, the RG approach completely
disregards the interference of the wave-function amplitudes between
different units at each RG step. For this reason it is not clear to
what extent this approach captures the main features and reproduces
the quantitative predictions at the QH transition.  Therefore, a
comparison of the RG results with the results of direct simulations of
the CC model is crucial. These direct simulations are usually carried
out by employing either the quasi-1D version \cite{MacK81} or the 2D
version \cite{FisL81} of the transfer-matrix method.  The results of
application of the version of Ref.\ \onlinecite{MacK81} to the CC
model are reported in Refs.\ \onlinecite{ChaC88} and
\onlinecite{LeeWK93}. In Refs.\ \onlinecite{WanJL96,ChoF97,JovW98} the
version of Ref.\ \onlinecite{FisL81} was utilized.  For the critical
exponent the values $\nu=2.5\pm0.5$ (Ref.\ \onlinecite{ChaC88}) and
later $\nu=2.4\pm0.2$ (Ref.\ \onlinecite{LeeWK93}) were obtained.
Note that our result is in excellent agreement with these values, and
is also consistent with the most precise $\nu=2.35\pm
0.03$.\cite{HucK90} This already indicates a remarkable accuracy of
the RG approach.  In Refs.\ \onlinecite{WanJL96,ChoF97,JovW98} the
critical distribution $P_{\text c}(G)$ of the conductance was studied.
$P_{\text c}(G)$ was found to be broad, which is in accordance with
Fig.\ \ref{fig-PQ}.  However, a more detailed comparison is
impossible, since the results of the simulations
\cite{WanJL96,ChoF97,JovW98} do not obey the electron-hole symmetry
condition $P_{\text c}(G)=P_{\text c}(1-G)$. On the other hand, within
the RG approach, the latter condition is satisfied automatically.
Nevertheless, we can compare the moments of $P_{\text c}(G)$ to those
calculated in Refs.\ \onlinecite{WanJL96} and \onlinecite{ChoF97}.
They are presented in Fig.\ \ref{fig-pmom}.  In Ref.\ 
\onlinecite{ChoF97} only the standard deviation $\left(\langle G^2
  \rangle - \langle G \rangle^2\right)^{1/2}\approx 0.3$ was computed.
Our result is $0.316$. In Ref.\ \onlinecite{WanJL96} the moments were
fitted by two analytical functions, which are also shown in Fig.\ 
\ref{fig-pmom}.  They agree with our calculations up to the sixth
moment. Here we point out that the moments obtained in Ref.\ 
\onlinecite{WanJL96} can hardly be distinguished from the moments of a
uniform distribution. This reflects the fact that $P_{\text c}(G)$ is
practically flat except for the peaks close to $G=0$ and $G=1$.

In Refs.\ \onlinecite{WanLS98} and  \onlinecite{AviBB99} $P_{\text c}(G)$ was studied by
methods which are not based on the CC model.  Both works reported a
broad distribution $P_{\text c}(G)$. In Ref.\  \onlinecite{WanLS98}
$P_{\text c}(G)$ was found to be almost flat. The major difference
between Ref.\ \onlinecite{WanLS98} and Fig.\ \ref{fig-PQ} is the
behavior of $P_{\text c}(G)$ near the points $G=0$ and $1$. That is,
$P(G)$ drops in Ref.\ \onlinecite{WanLS98} to zero at the ends, while
Fig.\ \ref{fig-PQ} exhibits maxima.  In Ref.\ \onlinecite{AviBB99}
the behavior of $P_{\text c}(G)$ is qualitatively similar to Fig.\
\ref{fig-PQ}, with maxima at $G=0$ and $1$.  However, the
statistics in Ref.\ \onlinecite{AviBB99} are rather poor, which again
rules out the possibility of a more detailed comparison with our
results.

Finally, we point out that our results agree completely with Refs.\ 
\onlinecite{WeyJ98,JanMW98,JanMMW98} where a similar RG treatment of
the CC model was carried out.  Our numerical data have a higher
resolution, and show significantly less statistical noise. This is
because we took advantage of faster computation by using the
analytical solution of the RG [Eq.\ (\ref{eq-qhrg})].\cite{GalR97}
Also, note that, in Refs.\ \onlinecite{WeyJ98} and
\onlinecite{JanMMW98} the critical exponent $\nu$ was calculated using
a procedure different from that described in Sec.\ \ref{ssec-RG-Q}.
Nevertheless, the values of $\nu$ determined by both methods are
close.
We emphasize that a systematic improvement of the RG procedure, i.e.,
by inclusion of more than five SP's into the basic RG unit as reported in
Refs.\ \onlinecite{WeyJ98,JanMW98,JanMMW98}, leads to similar results.
The critical distribution of the conductance was also studied
experimentally \cite{CobK96,CobBF99} in mesoscopic QH samples.
Although an almost uniform conductance distribution consistent with
theoretical predictions was found in Ref.\ \onlinecite{CobK96}, further
detailed analysis of the mesoscopic pattern \cite{CobBF99} has
revealed the crucial role of the charging effects, which were
neglected in all theoretical studies.

We conclude that the test of the RG approach against other simulations
proves that this approach provides a very accurate quantitative
description of the QH transition.  It can now be used to study the
influence of long-ranged correlations.

\section{Macroscopic inhomogeneities}
\label{sec-inh}
\subsection{General considerations}
\label{ssec-inh-des}
A natural way to incorporate a quenched disorder into the CC model is
to ascribe a certain random shift, $z_Q$, to each SP height, and to
assume that the shifts at different SP positions, ${\bf r}$ and ${\bf
  r}'$, are correlated as
\begin{equation}
\label{eq-corr}
\langle z_Q({\bf r})z_Q({\bf r}')\rangle
\propto |{\bf r}-{\bf r}'|^{-\alpha},
\end{equation}
with $\alpha >0$. After this, the conventional transfer-matrix methods
of Refs.\ \onlinecite{MacK81} and \onlinecite{FisL81} could be
employed for numerically precise determination of $\langle G \rangle$,
the distribution $P_{\text c}(G)$, its moments, and most importantly,
the critical exponent, $\nu$.  However, the transfer-matrix approach
for a 2D sample is usually limited to fairly small sizes (e.g., up to
$128$ in Ref.\ \onlinecite{JovW98}) due to the numerical complexity of
the calculation. Therefore, the spatial decay of the power-law
correlation by, say, more than an order in magnitude is hard to
investigate for small $\alpha$.  In principle the quasi-1D
transfer-matrix method \cite{ChaC88,MacK81} can easily handle such
decays at least in the longitudinal direction, where typically a few
million lattice sites are considered iteratively. A major drawback,
however, is the numerical generation of power-law correlated
randomness since no iterative algorithm is known.\cite{Pra92,MakHSS96}
This necessitates the complete storage of different samples of
correlated SP height landscapes,\cite{RusKBH99} and the advantage of
the iterative transfer-matrix approach is lost.  Furthermore, in order
to deduce the critical exponent,\cite{LeeWK93} one needs to perform
finite-size scaling \cite{MacK81} with transverse sizes that should
also be large enough to capture the main effect of the power-law
disorder in the transverse direction.  Consequently, even for a single
transfer-matrix sample, the memory requirements add up to gigabytes.

On the other hand, the RG approach is perfectly suited to study the
role of the quenched disorder.  First, after each step of the RG
procedure, the effective system size doubles.  At the same time, the
magnitude of the smooth potential, corresponding to the spatial scale
$r$, falls off with $r$ as $r^{-\alpha/2}$. As a result, the
modification of the RG procedure due to the presence of the quenched
disorder reduces to a rescaling of the disorder magnitude by a {\em
  constant} factor $2^{-\alpha/2}$ at each RG step.  Second, the RG
approach operates with the conductance distribution, $P_n(G)$, which
carries information about {\em all} the realizations of the quenched
disorder within a sample of a size $2^n$.  This is in contrast to the
transfer-matrix approach,\cite{MacK81,FisL81} within which a
small increase of the system size requires the averaging over the
quenched disorder realizations to be conducted again.

In order to find out whether or not the critical behavior
is affected by the quenched disorder, the following argument was
put forward in Ref.\ \onlinecite{WeiH83}. In the absence of the quenched disorder,
the correlation length, $\xi_0$, for a given energy $z_Q$ in the
vicinity of the transition is proportional to $z_Q^{-\nu}$.
Now consider a sample with an area $A=\xi_0^2$. The variance of
the quenched disorder within the sample is given by
\begin{equation}
\label{eq-square}
\begin{array}{lll}
\Delta_0^2 & = & \frac{1}{A^2}\left\langle\int_A d^2r z_Q({\bf r})\int_A d^2
r' z_Q({\bf r'})\right\rangle \\
& = & \frac{1}{A^2}\int_A d^2r \int_A d^2r'
\langle z_Q({\bf r})z_Q({\bf r'})\rangle \\
& \propto & \xi_0^{-2}\int_0^{\xi_0} dr r^{1-\alpha},
\end{array}
\end{equation}
where the  last relation follows  from Eq.\ (\ref{eq-corr}).

The critical behavior remains unaffected by the quenched disorder if
the condition $\Delta_0^2/z_Q^2 \rightarrow 0$
as $z_Q\rightarrow 0$ is met. Using Eq. (\ref{eq-square}), the
ratio $\Delta_0^2/z_Q^2$ can be presented as
\begin{equation}
\label{eq-ratio}
\frac{\Delta_0^2}{z_Q^2}\propto \left \{
\begin{array}{ll}
z_Q^{2\nu -2},                 & \alpha>2 \quad ,\\
z_Q^{2\nu -2}\ln (z_Q^{-\nu}), & \alpha=2 \quad ,\\
z_Q^{\alpha\nu -2},            & \alpha<2 \quad .
\end{array}\right.
\end{equation}
We thus conclude that quenched disorder is irrelevant when
\begin{equation}
\begin{array}{ll}
\nu > 1,        & {\text{for }} \alpha \geq 2 , \\
\alpha \nu > 2, & {\text{for }}\alpha <2 \quad .
\end{array}
\label{eq-extended-harris}
\end{equation}
The first condition corresponds to the original Harris criterion
\cite{Har74} for  uncorrelated disorder, while the second condition is
the extended Harris criterion.\cite{WeiH83} It yields the critical value of the
exponent $\alpha$, i.e., $\alpha_c=2/\nu$.

The above consideration suggests the following algorithm. For the
homogeneous case all SP's constituting the new super-SP are assumed to
be identical, which means that the distribution, of heights, $Q_n(z)$,
for all of them is the same.  For the correlated case these SP's are no
longer identical, but rather their heights are randomly shifted by the
long-ranged potential.  In order to incorporate this potential into
the RG scheme, $Q_n(z_i)$ should be replaced by
$Q_n(z_i-\Delta^{(n)}_i)$ for each SP, $i$, where $\Delta^{(n)}_i$ is
the random shift. Then the power-law correlation of the quenched
disorder enters into the RG procedure through the distribution of
$\Delta^{(n)}_i$. That is, for each $n$ the distribution is Gaussian
with the width $\beta (2^n)^{-\alpha/2}$.  For large enough $n$ the
critical behavior should not depend on the magnitude $\beta$, but on
the power, $\alpha$, only.
%
\subsection{Numerical results}
\label{ssec-inh-res}
Here we report the results of the application of the algorithm
outlined in the previous section.  First, we find that for all
values of $\alpha>0$ in correlator (\ref{eq-corr}) the FP
distribution is identical to the uncorrelated case within the accuracy
of our calculation.  In particular, $\langle G \rangle=0.498$ is
unchanged.  However, the convergence to the FP is numerically less
stable than for uncorrelated disorder due to the correlation-induced
broadening of $Q_n(z)$ during each iteration step.
In order to compute the critical exponent $\nu(\alpha)$ we start the
RG procedure from $Q_0(z-z_0)$, as in the uncorrelated case, but, in
addition, we incorporate the random shifts caused by the quenched
disorder in generating the distribution of $z$ at each RG step.  The
results shown in Fig.\ \ref{fig-nu_b} illustrate that for increasing
long-ranged character of the correlation (decreasing $\alpha$) the
convergence to a limiting value of $\nu$ slows down drastically.  Even
after eight RG steps (i.e., a magnification of the system size by a factor
of $256$), the value of $\nu$ with long-ranged correlations still
differs appreciably from $\nu=2.39$ obtained for the uncorrelated
case.  The RG procedure becomes unstable after eight iterations, i.e.,
$z_{\text{max},9}$ can no longer be obtained reliably from $Q_9(z)$.
Unfortunately, for small $\alpha$ the convergence is too slow to yield
the limiting value of $\nu$ after eight steps only.  For this reason, we
are, strictly speaking, unable to unambiguously answer the question
whether sufficiently long-ranged correlations result in an
$\alpha$-dependent critical exponent $\nu(\alpha)$, or the value of
$\nu$ eventually approaches the uncorrelated value of $2.39$.
Nevertheless, the results in Fig.\ \ref{fig-nu_b} indicate that the
effective critical exponent exhibits a sensitivity to the long-ranged
correlations even after a large magnification by $256\times 256$.
Therefore, in realistic samples of finite sizes, macroscopic
inhomogeneities are able to affect the results of scaling studies.
Note further that as shown in the inset of Fig.\ \ref{fig-nu_b} there
is no simple scaling of $\nu$ values when plotted as function of an
renormalized system size $2^{\alpha n/2}$.
We emphasize that $\nu(\alpha)$ obtained after eight RG steps always
{\em exceeds} the uncorrelated value.  Thus, our results indicate that
macroscopic inhomogeneities must lead to smaller values of
$\kappa\propto 1/\nu$.  Experimentally, the value of $\kappa$ smaller
than $0.42$ was reported in a number of early (see, e.g., Ref.\ 
\onlinecite{KocHKP92} and references therein) as well as recent
\cite{Lan00} works. This fact was accounted for by different reasons
(such as temperature dependence of the phase breaking time, incomplete
spin resolution, valley degeneracy in Si-based metal-oxide-semiconductor 
field-effect transistors, and
inhomogeneity of the carrier concentration in GaAs-based structures
with a wide spacer).  Briefly, the spread of the $\kappa$ values was
attributed to the fact that the temperatures were not low enough to
assess the truly critical regime.  The possibility of having $\kappa <
0.4$ due to the correlation-induced dependence of the effective $\nu$ on
the phase-breaking length or, ultimately, on the sample size, as in
Fig.\ \ref{fig-nu_b}, was never considered.

\begin{figure}[p]
\centerline{\includegraphics[width=0.95\columnwidth]{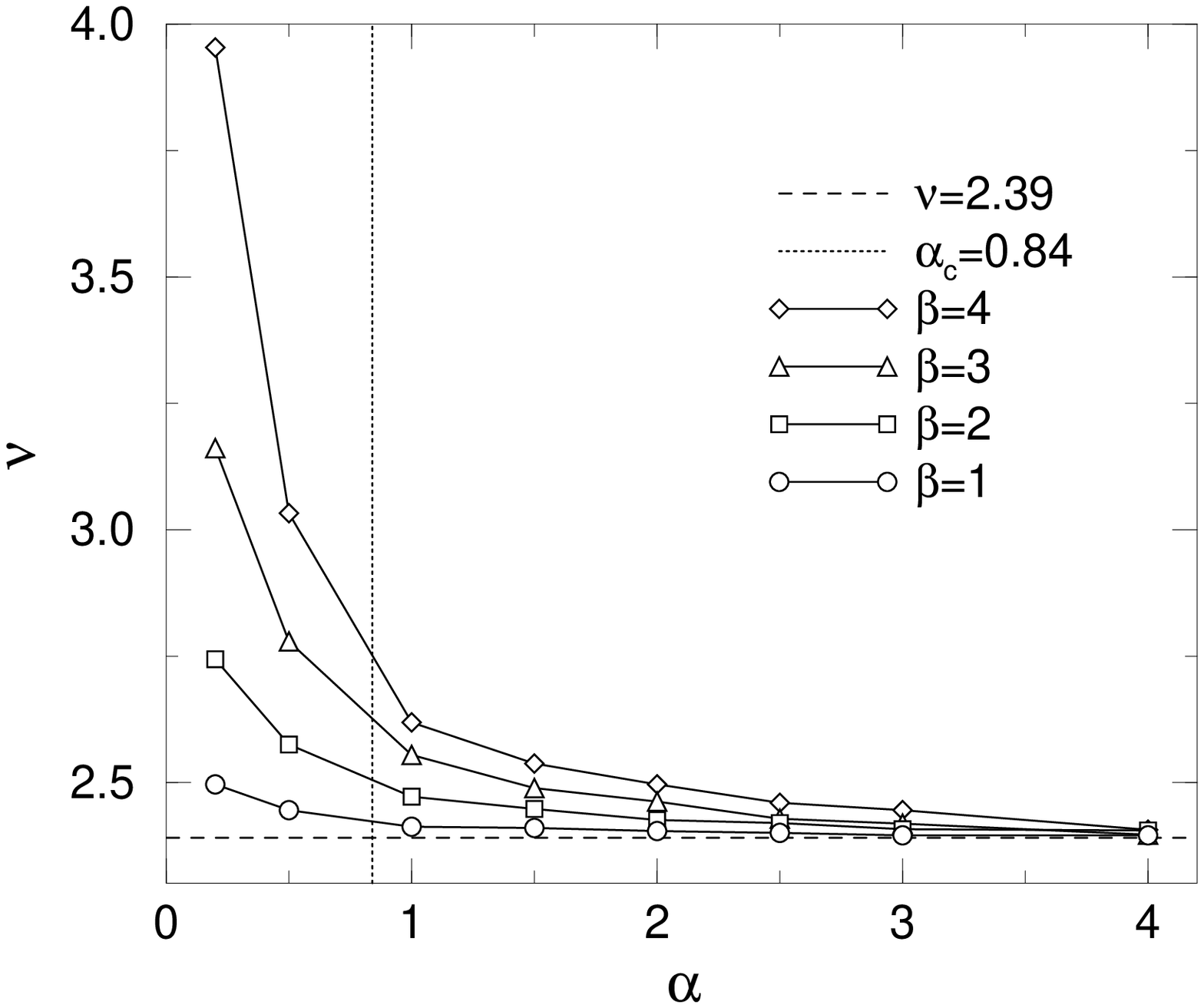}}
\caption{\label{fig-alpha}
  Dependence of the critical exponent $\nu$ on correlation exponent
  $\alpha$ for different $\beta=1, 2, 3$, and $4$ as obtained after
  eight QH-RG iterations.  The dotted line indicates $\alpha_c=0.84$,
  following from the extended Harris criterion (Ref.\ 
  \protect\onlinecite{WeiH83}) for classical percolation.}
\end{figure}

Figure \ref{fig-alpha} shows the values of $\nu$ obtained after the
eighth RG step as a function of the correlation exponent $\alpha$ for
different dimensionless strengths $\beta$ of the quenched disorder.
It is seen that in the domain of $\alpha$, where the values of $\nu$
differ noticeably from $\nu=2.39$, their dependence on $\beta$ is
strong.  According to the extended Harris criterion, $\nu(\alpha)$ is
expected to exhibit a cusp at the $\beta$-independent value
$\alpha=\alpha_c=2/2.39 \approx 0.84$.  From our results in Fig.\ 
\ref{fig-alpha}, two basic observations can be made. For a small enough
magnitude of the long-range disorder, we see a smooth enhancement of
$\nu(\alpha)$ with decreasing $\alpha$ without a cusp. Although such a
behavior is due to the relatively small number of RG steps, the data
might be relevant for realistic samples which have a finite size and
a finite phase-breaking length governed by temperature. At the largest
$\beta=4$, the cusp eventually shows up but the numerics becomes
progressively ambiguous, forbidding us from going to even larger $\beta$.
The origin for this strong $\beta$ dependence of our results is a
profound difference between the classical and quantum percolation
problems. This difference is discussed in the Sec. \ref{sec-sum}.

\section{Discussion}
\label{sec-sum}

\subsection{Classical case}
\label{subsec-classical}
In the classical limit, the motion of an a electron in a strong magnetic
field and a smooth potential reduces to the drift of the Larmour circle
along the equipotential lines.  Correspondingly, the description of the
delocalization transition reduces to the classical percolation problem.
As mentioned above, for classical percolation the quenched
disorder is expected to cause a crossover in the exponent $\nu_c$,
describing the size of a critical equipotential from $\nu_c=4/3$ to
$\nu_c=2/\alpha$ for $\alpha < 3/2$.  This prediction\cite{Wei84} was
made on the basis of Eq.\ (\ref{eq-square}). It was later tested by
numerical simulations\cite{Pra92} which utilized the Fourier filtering
method to generate a long-range-correlated random potential.  The
exponent $\nu_c(\alpha)$ was studied using the same classical real-space
RG method \cite{ReyKS77} that we utilized above. The values of $\nu_c$
inferred for $\alpha > 1$ were consistently lower than $2/\alpha$.  For
example, $\nu_c=5$ was found for $\alpha=0.4$,\cite{Wei84} whereas
$\nu_c=3.4\pm 0.3$ was observed.

The classical version of the delocalization transition is instructive,
since it allows one to trace how the critical equipotentials grow in
size upon approaching the percolation threshold, and how the quenched
disorder affects this growth. Roughly speaking, in the absence of
long-ranged correlations, the growth of the equipotential size is due
to the attachment of smaller equipotentials to the critical ones.  As
a result, the shape of a critical equipotential is dendritelike. As
the threshold is approached, different critical equipotentials become
connected through the narrow ``arms'' of the dendrite.  Long-ranged
correlations change this scenario drastically.  As could be expected
intuitively, and as follows from the simulations,\cite{MakHSS96}
critical equipotentials become more compact due to correlations.  In
fact, for $\alpha < 0.25$, the ``arms'' play no role,\cite{Pra92}
i.e., the morphology of a critical equipotential becomes identical to
its ``backbone''.  As a result, the formation of the infinite
equipotential at the threshold occurs through a sequence of ``broad''
merges of compact critical equipotentials.  The correlation-induced
enhancement of $\nu_c$ indicates that due to these merges the size of
the critical equipotential in the close vicinity of the threshold
grows faster than in the uncorrelated case.
Since our simulations also demonstrate the enhancement of the
critical exponent due to the correlations, the main result
of the present paper can be formulated as follows: quenched disorder
affects classical and quantum percolation in a similar fashion.

\subsection{Quantum case}
\label{subsec-quantum}

Here we note that there is a crucial distinction between the classical
case and the quantum regime of the electron motion considered in the
present paper. Indeed, within the classical picture, correlated disorder
implies that the motion of the guiding centers of the Larmour orbits in
two neighboring regions is completely identical.  In our study, we have
incorporated the correlation of {\em heights} of the saddle points into
the RG scheme. At the same time we have assumed that the {\em
Aharonov-Bohm phases} acquired by an electron upon traversing the
neighboring loops are completely {\em uncorrelated}. This assumption
implies that, in addition to the long-ranged potential, a certain
short-ranged disorder causing a spread in the perimeters of neighboring
loops of the order of the magnetic length is present in the sample. The
consequence of this short-range disorder is the sensitivity of our
results to the value of $\beta$ which parametrizes the magnitude of the
correlated potential. The presence of this short-range disorder
affecting exclusively the Aharonov-Bohm phases significantly complicates 
the observation of a cusp in the $\nu (\alpha)$ dependence at $\alpha
\approx 0.84$, as might be expected from the extended Harris criterion.

Let us elaborate on these complications. A general form of the
correlator for long-range disorder is
\begin{equation}
  \label{eq-correlator}
   \langle V({\bf r}) V({\bf r'}) \rangle =
   V_0^2 F\left(\frac{|{\bf r}-{\bf r}'|}{l}\right)
\end{equation}
where $l$ is the microscopic length; $F(0)=1$, $F(x)\propto x^{-\alpha}$
for $x \gg 1$. Now suppose that the correlator contains an additional
short-range term $W_0^2 G(|{\bf r}-{\bf r}'|/l)$, with $G(0)=1$ and
$G(x)$ falling off much more rapidly than $F(x)$ for $x \gg 1$. The
extended Harris criterion implies that this term will not change
$\nu(\alpha)$ in an {\em infinite sample}. It is obvious, however, that
in order to ``erase the memory'' of short-range disorder, many more
RG iterations have to be performed or, equivalently, much larger system
sizes should be analyzed. Moreover, the larger  the ratio $W_0 / V_0$,
the more challenging the numerics becomes.
At this point, we emphasize that in quantum percolation the short-range
term emulated by the randomness in the phases has a huge magnitude.
Indeed, if we choose in the first RG step all five SP's to be identical
with power transmission coefficients equal to $0.5$, then due to the
phases, the width of the $Q(z)$ distributions after the first step is
already $\pm 2.5$.\cite{GalR97} This translates into an enormously wide
spread in the transmission coefficients of effective SP's ranging from
$0.075$ to $0.92$. In order to suppress this intrinsic ``quantum white
noise'', one either has to perform more RG steps or to increase the
magnitude of $V_0$ (parameter $\beta$ in the notation of Sec.\
\ref{sec-inh}). Both strategies are limited by numerical instabilities.
In particular, a larger $\beta$ leads to more weight in the tails of the
$Q(z)$ distribution in which the uncertainty is maximal. In other words,
at large $\beta$ the role of rare realizations is drastically
emphasized.

In fact, if we had to draw a quantitative conclusion on the basis of the
accuracy we have achieved, we would base it on the curve in Fig.\
\ref{fig-alpha} corresponding to the maximal value $\beta=4$. Actually,
for this $\beta$, the agreement with the extended Harris criterion is
fairly good. In particular, for $\alpha=0.5$, we find $\nu\approx 3$,
whereas $2/\alpha = 4$.

We also want to point out that the limited number (eight) of RG steps
permitted by the numerics nevertheless allows us to trace the
evolution of the wave functions from {\em microscopic} scales (of the
order of the magnetic length) to {\em macroscopic} scales (of the
order of $5 \mu{\rm m}$) which are comparable to the sizes of the
samples used in the experimental studies of scaling (see e.g., Refs.\ 
\onlinecite{KocHKP91b} and \onlinecite{KocHKP92}) and much larger than
the samples\cite{CobK96,CobBF99} used for the
studies of mesoscopic fluctuations.

\subsection{Concluding remarks}
\label{subsec-conclude}

It was argued for a long time that the enhanced value of the critical
exponent $\nu$ extracted from the narrowing of the transition region
with temperature has its origin in the long-ranged disorder present in
GaAs-based samples. To our knowledge, the present work is the first attempt to quantify
this argument. We indeed find that the random potential with a
power-law correlator leads to the values of $\nu$ exceeding $\nu
\approx 2.35$, which is firmly established for short-ranged disorder.
Another {\em qualitative} conclusion of our study is that the spatial
scale at which the exponent $\nu$ assumes its ``infinite-sample''
value is much larger in the presence of the quenched disorder than in
the uncorrelated case. In fact this scale can be of the order of
microns.  This conclusion can also have serious experimental
implications. That is, even if the sample size is much larger than this
characteristic scale, this scale can still exceed the phase-breaking
length, which would mask the true critical behavior at the QH
transition.

\begin{figure}[p]
\centerline{\includegraphics[width=0.95\columnwidth]{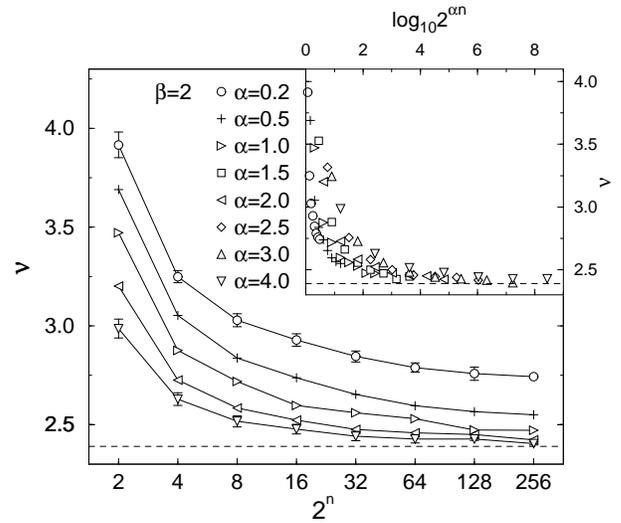}}
\caption{\label{fig-nu_b}
  Critical exponent $\nu$ obtained by the QH-RG approach as a function
  of RG scale $2^n$ for $\beta=2$ and different correlation exponents
  $\alpha$.  The dashed line indicates $\nu=2.39$, which is the value
  that we obtain for uncorrelated disorder. For clarity, we show the
  errors only for $\alpha=0.2$ and $4$.  Inset: $\nu$ vs $2^{\alpha
    n/2}$ does not scale for, e.g., $\beta=2$.  }
\end{figure}

Our numerical results demonstrate that when the critical exponent
depends weakly on the sample size (large $n$ in Fig.\ \ref{fig-nu_b}),
the ``saturated'' value of $\nu$ depends crucially on the
``strength'' $\beta$ of the quenched disorder. Thus it is important
to relate this strength to the observable quantities. We can roughly
estimate $\beta$ assuming that themicroscopic spatial scale (lattice
constant) is the magnetic length, $l$, while the microscopic energy
scale (the SP height) is the width of the Landau level.  We denote by
$\gamma$ a typical fluctuation of the filling factor within a region
with size $L$.  Then the estimate for $\beta$ is $\beta \sim
\gamma \left(L/l\right)^{\alpha/2}$.  Naturally, for a given $\gamma$,
the larger values of $\alpha$ correspond to the ``stronger'' quenched
disorder parameter $\beta$.

Note, finally, that throughout this paper we have considered the
localization of a single electron. The role of electron-electron
interactions in the scaling of the integer QH effect was recently
addressed in Refs.\ \onlinecite{HucB99} and  \onlinecite{WanFGC00}.
\acknowledgements
We thank F.\ Evers, F.\ Hohls, R.\ Klesse, G.\ Landwehr, A.\ Mirlin,
and A.\ Zee for stimulating discussions.  This work was supported by
the NSF-DAAD collaborative research Grant No. INT-0003710.  P.C., R.A.R., and
M.S. also gratefully acknowledge the support of DFG within the
Schwer\-punkt\-pro\-gramm ``Quanten-Hall-Systeme'' and SFB~393.


\begin{thebibliography}{10}
\bibitem{inhomo} We emphasize that sometimes the term "long-ranged
  disorder" is also used for a disorder that has a finite correlation
  radius but which is larger than the magnetic length, see, e.g., F.\
  Evers, A.~D. Mirlin, D.~G. Polyakov, and P. W\"olfle, Phys.\ Rev.\ B
  {\bf 60}, 8951 (1999). This is different from the present situation.
\end{thebibliography}
\end{document}